\documentclass[fleqn,usenatbib]{mnras}
\usepackage{newtxtext,newtxmath}
\usepackage{graphicx}
\usepackage{amsmath}
\usepackage{booktabs}
\usepackage{microtype}

\title[Vortex pinning and crustal elasticity]{Vortex pinning and the elastic response of neutron-star crusts -- I. Axisymmetric loading}
\author[E. Giliberti \& G. Cambiotti]{
E. Giliberti$^{1}$\thanks{E-mail: eliagiliberti@gmail.com}
and G. Cambiotti$^{2}$\\
$^{1}$Istituto Leonardo da Vinci, Cologno Monzese, Italy\\
$^{2}$Dipartimento di Scienze della Terra ``Ardito Desio'', Universit\`a degli Studi di Milano, Milano, Italy
}
\date{30 August 2026}
\pubyear{2026}

\begin{document}
\label{firstpage}
\pagerange{\pageref{firstpage}--\pageref{lastpage}}
\maketitle

\begin{abstract}
Vortex pinning is central to the standard interpretation of pulsar glitches, but the mechanical load exerted by a pinned vortex array on the solid crust has received less attention than the angular-momentum reservoir itself. We calculate the axisymmetric elastic response of a continuously stratified neutron-star crust to this load using the elastic--gravitational framework of \citet{Giliberti2019,Giliberti2020}, SLy4 and BSk21 stellar backgrounds, and composition-dependent Coulomb shear moduli. The superfluid--crust lag sets the Magnus force per unit vortex length, while the mesoscopic pinning force of \citet{Seveso2016} provides a local upper bound. At low lag the response is linear; progressive local saturation then produces a broad transition and a finite high-lag envelope. The stress maximum is robustly located at the deep crustal boundary towards the rotation axis, although the local Magnus force vanishes on-axis, showing that the dominant localization is produced by global elastic--gravitational redistribution rather than by the local forcing amplitude. Composition-dependent elasticity changes the stress amplitude by 10--20 per cent relative to the common $\mu=10^{-2}P$ prescription and, importantly, reverses the SLy4--BSk21 ordering, demonstrating that the simplified modulus is not a universal rescaling. At a Vela-motivated benchmark $\Delta\Omega=10^{-2}\,{\rm rad\,s^{-1}}$, the maximum strain is only $3.1\times10^{-5}$ (SLy4) and $4.7\times10^{-5}$ (BSk21), and remains below $1.4\times10^{-4}$ on the formal plateau. A $1.2$--$2.0\,M_\odot$ scan changes the stress amplitude by only about 20 per cent and leaves the deep-polar localization unchanged. The same Vela-scale lag contains enough differential angular momentum to account for the 2016 glitch provided a reservoir of order one per cent of the stellar moment of inertia participates efficiently. Pinning therefore supplies a structured and astrophysically relevant crustal pre-stress, but cannot by itself break an initially relaxed crust.
\end{abstract}


\section{Introduction}
Pulsar glitches show that a neutron star can sustain differential rotation between dynamically distinct components. In the standard picture, neutron-superfluid rotation is carried by quantized vortices. As the electromagnetically coupled component spins down, outward vortex motion is required for the superfluid to follow the secular decrease in angular velocity. Pinning of vortices to the nuclear structures of the inner crust can impede this motion, allowing a rotational lag and an angular-momentum reservoir to develop before a glitch \citep{AndersonItoh1975,Alpar1984,HaskellPizzocheroSidery2012,HaskellMelatos2015,AntonelliPizzochero2017}.

The same physics has a direct mechanical consequence. A pinned vortex subject to the Magnus force transfers the opposite reaction force to the lattice. An array of pinned vortices therefore loads the solid crust continuously while the lag is being built. Microscopic vortex--nucleus calculations have evolved from semi-classical treatments \citep{DonatiPizzochero2004,DonatiPizzochero2006} to time-dependent density-functional and Hartree--Fock--Bogoliubov approaches \citep{Wlazlowski2016,Klausner2023}. For the stellar-scale problem, however, the relevant quantity is not the force at a single nucleus but the mesoscopic force per unit length of an extended vortex sampling many lattice sites. We adopt the density-dependent maximum force calculated by \citet{Seveso2016}.

The purpose of this paper is to connect that pinning microphysics to the global elastic response of a realistic stratified crust. The mechanical framework follows the sequence developed in \citet{Giliberti2019} and generalized to continuously stratified compressible stars by \citet{Giliberti2020}. This sits within a broader literature on global elastic deformations of neutron-star crusts, from the stress-integral approach of \citet{Ushomirsky2000} and non-Cowling boundary-value calculations \citep{HaskellJonesAndersson2006,JohnsonMcDanielOwen2013} to more recent force-based formulations \citep{GittinsAndersson2021,MoralesHorowitz2022}. The present problem replaces the rotational body force considered in those works by the reaction force of the pinned vortex array. The physical chain is therefore
\begin{equation}
 \Delta\Omega \longrightarrow f_{\rm M}\longrightarrow f_{\rm line}
 \longrightarrow f_{\rm vol}\longrightarrow \{\boldsymbol\xi,u_{ij},\sigma_{ij}\}.
\end{equation}
The distinction is important: a lag is not itself a crustal stress. The stress follows only after the vortex force is transmitted to the lattice and redistributed by the elastic--gravitational response of the whole crust.

We use the label Paper I to mark the scope of the present calculation: here we establish the axisymmetric elastostatic baseline and ask three questions: how the stress grows as lag accumulates, where the crust carries the load, and whether the resulting strain can approach failure. Non-axisymmetric mountains, time-dependent unpinning and glitch dynamics are planned extensions rather than ingredients required by the present results.

\section{Elastic--gravitational response}
\subsection{The Giliberti crust model}
We consider a Newtonian spherical star in hydrostatic equilibrium,
\begin{equation}
 \frac{dP}{dr}=-\rho g,\qquad g=\frac{Gm(r)}{r^2},\qquad
 \frac{dm}{dr}=4\pi r^2\rho .
\end{equation}
The fluid core is surrounded by an elastic crust. The reference state is mechanically relaxed and the perturbation is treated in linear elasticity. We use the continuously stratified formalism of \citet{Giliberti2020}, which extends the analytical incompressible treatment of \citet{Giliberti2019} by retaining realistic radial stratification, compressibility and the perturbation of the gravitational potential.

For an axisymmetric spheroidal displacement,
\begin{equation}
\boldsymbol\xi=\sum_\ell\left[U_\ell(r)Y_{\ell0}\,\boldsymbol e_r
+V_\ell(r)\nabla_\perp Y_{\ell0}\right],
\end{equation}
the perturbed force balance and Poisson equations are
\begin{align}
 \nabla_j\delta\sigma^{ij}-\delta\rho\nabla^i\Phi
 -\rho\nabla^i\delta\Phi+f_{\rm ext}^{\,i}&=0,\\
 \nabla^2\delta\Phi&=4\pi G\delta\rho .
\end{align}
The elastic stress is
\begin{equation}
 \delta\sigma_{ij}=-\delta P\,\delta_{ij}
 +2\mu\left(u_{ij}-\frac{1}{3}u^k_{\ k}\delta_{ij}\right),
\end{equation}
where $u_{ij}=(\nabla_i\xi_j+\nabla_j\xi_i)/2$. At the crust--core interface we impose the fluid--solid matching conditions of \citet{Giliberti2020}; at the outer elastic boundary the traction vanishes and the gravitational perturbation is matched to the exterior multipole. The gravitational matching variable is
\begin{equation}
 q=\psi'+\frac{\ell+1}{r}\psi-\frac{4\pi G\rho}{g}\psi .
\end{equation}

\subsection{Realistic shear modulus and strain diagnostic}
The production calculations use a composition-dependent Coulomb shear modulus,
\begin{equation}
 \mu=0.1194\,n_i\frac{(Ze)^2}{a},\qquad
 a=\left(\frac{3}{4\pi n_i}\right)^{1/3},
 \label{eq:mu}
\end{equation}
appropriate to an angle-averaged body-centred Coulomb lattice \citep{Strohmayer1991}. For SLy4, $Z$ and the Wigner--Seitz cell radius are taken from the inner-crust composition of \citet{DouchinHaensel2001}; for BSk21 we use the composition of \citet{Pearson2012} and its analytic representation by \citet{Potekhin2013}. The corresponding crust--core transitions are treated consistently with each EoS. A comparison with the commonly used control prescription $\mu=10^{-2}P$ is given in Appendix~\ref{app:numerics}.

Let $\lambda_{\max}$ and $\lambda_{\min}$ be the extreme principal strains. We define
\begin{equation}
 \alpha=\lambda_{\max}-\lambda_{\min},\qquad
 \tau_{\rm T}=\frac{\sigma_{\max}-\sigma_{\min}}{2}=\mu\alpha .
 \label{eq:tresca}
\end{equation}
The stress and strain maxima need not coincide in a stratified crust. We use $\tau_{\rm T,max}$ to identify where the load is carried and $\alpha_{\max}$ to assess proximity to failure.

\section{The pinning force}
\subsection{Magnus loading and the mesoscopic cap}
Let
\begin{equation}
 \Delta\Omega=\Omega_{\rm s}-\Omega_{\rm c}
\end{equation}
be the superfluid--crust lag. For straight vortices parallel to the rotation axis, the Magnus force per unit vortex length is
\begin{equation}
 f_{\rm M}=\rho_{\rm s}\kappa\varpi|\Delta\Omega|
 =\rho_{\rm s}\kappa r\sin\theta|\Delta\Omega|,
 \label{eq:magnus}
\end{equation}
where $\kappa=h/(2m_n)=1.99\times10^{-3}\,{\rm cm^2\,s^{-1}}$. The vortex areal density is $n_v=2\Omega_{\rm s}/\kappa$. We adopt $\Omega=70.4\,{\rm rad\,s^{-1}}$, as the rounded reference value used in the production solver, appropriate to Vela's spin period $P\simeq89$ ms \citep{Manchester2005}. The event-specific value $\Omega\simeq70.6\,{\rm rad\,s^{-1}}$ used in Section~6.2 differs by only $0.3$ per cent; we retain the production value here rather than renormalizing the computed stress field a posteriori.

\begin{figure}
\centering
\includegraphics[width=\columnwidth]{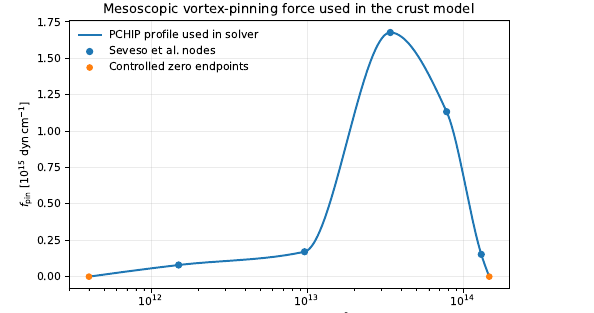}
\caption{Mesoscopic maximum pinning force per unit vortex length from \citet{Seveso2016}. The local Magnus force is allowed to grow only up to this density-dependent cap.}
\label{fig:fpin}
\end{figure}

The lattice cannot transmit an arbitrarily large force while remaining pinned. We therefore impose the mesoscopic pinning limit of \citet{Seveso2016},
\begin{equation}
 f_{\rm line}(r,\theta)=\min\left[f_{\rm M}(r,\theta),
 f_{\rm pin}(\rho)\right].
 \label{eq:hardmin}
\end{equation}
The adopted profile is shown in Fig.~\ref{fig:fpin}. It is interpolated in $\log\rho$ and brought continuously to zero at the EoS-specific inner boundary of the pinning region. The taper is physically motivated by the disappearance of the crustal nuclear lattice, and therefore of the crustal pinning mechanism represented by the adopted profile, on entering the fluid core; its detailed shape is not fixed by the mesoscopic calculation and also provides a numerically smooth endpoint. Alternative endpoint treatments change the final stress only at the few-per-cent level (Appendix~\ref{app:numerics}).

The Seveso profile is used here as a definite microphysical input, not as an uncertainty-free measurement of the pinning force. Vortex rigidity, pairing, lattice geometry and the microscopic vortex--nucleus interaction all affect the inferred pinning energy and its conversion to a force per unit length \citep{DonatiPizzochero2004,DonatiPizzochero2006,Wlazlowski2016,Klausner2023}. Recent microscopic calculations still span roughly $10^{14}$--$10^{16}\,\mathrm{dyn\,cm^{-1}}$ in relevant crustal conditions \citep{Klausner2023}. The endpoint test in Appendix~\ref{app:numerics} therefore quantifies only the numerical/systematic sensitivity to how the adopted Seveso curve is terminated, not the full microphysical uncertainty in vortex pinning.

\subsection{From vortices to a stellar body force}
Coarse-graining over the vortex array gives
\begin{equation}
 f_{\rm vol}=n_v f_{\rm line}.
\end{equation}
The reaction force is directed along the cylindrical radial direction,
\begin{equation}
 \boldsymbol e_\varpi=\sin\theta\,\boldsymbol e_r+
 \cos\theta\,\boldsymbol e_\theta,
\end{equation}
so $f_r=f_{\rm vol}\sin\theta$ and $f_\theta=f_{\rm vol}\cos\theta$. The force is projected onto even axisymmetric spherical harmonics only after the local cap in equation~(\ref{eq:hardmin}) has been applied. The explicit projection is given in Appendix~\ref{app:projection}.

\section{From linear loading to pinning saturation}
\begin{figure}
\centering
\includegraphics[width=\columnwidth]{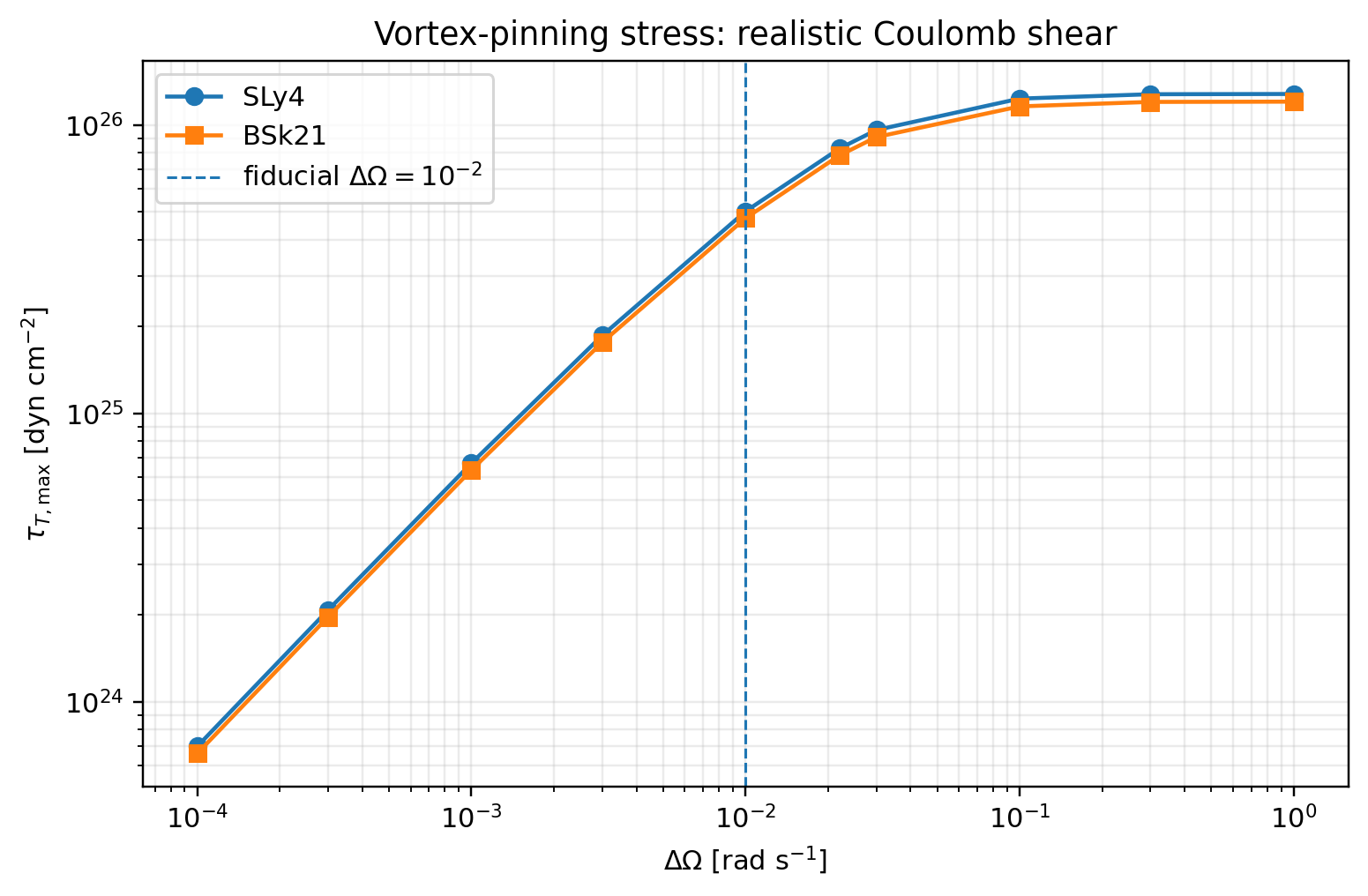}
\caption{Maximum Tresca stress versus superfluid--crust lag for $1.4\,M_\odot$ models with EoS-consistent Coulomb shear moduli. The vertical dashed line marks the fiducial $\Delta\Omega=10^{-2}\,{\rm rad\,s^{-1}}$, which lies in the partially saturated transition rather than in the purely linear regime.}
\label{fig:lag}
\end{figure}

Figure~\ref{fig:lag} shows the production lag scan for $1.4\,M_\odot$ SLy4 and BSk21 models, using equation~(\ref{eq:mu}) and harmonics through $\ell_{\max}=20$. At small lag the entire pinned region is Magnus limited, so $f_{\rm vol}\propto\Delta\Omega$ and linear elasticity gives
\begin{equation}
 \tau_{\rm T,max}\propto|\Delta\Omega|.
\end{equation}
The numerical solution follows this scaling closely below a few $10^{-3}\,{\rm rad\,s^{-1}}$.

At larger lag, different parts of the inner crust reach $f_{\rm pin}(\rho)$ at different times. Saturation is therefore progressive rather than a single global event. At the fiducial $\Delta\Omega=10^{-2}\,{\rm rad\,s^{-1}}$ we find
\begin{align}
 \tau_{\rm T,max}^{\rm SLy4}&=5.01\times10^{25}\,{\rm dyn\,cm^{-2}},\\
 \tau_{\rm T,max}^{\rm BSk21}&=4.75\times10^{25}\,{\rm dyn\,cm^{-2}}.
\end{align}
These values are already about 28 per cent below a linear extrapolation from the low-lag regime, while they are only about 39 per cent of the final plateau. Thus $10^{-2}\,{\rm rad\,s^{-1}}$ lies unambiguously in a \emph{partially saturated transition regime}.

The curves reach more than 96 per cent of their asymptotic value by $\Delta\Omega=0.1\,{\rm rad\,s^{-1}}$ and are essentially saturated by $0.3\,{\rm rad\,s^{-1}}$. The plateaux are $1.28\times10^{26}$ and $1.20\times10^{26}\,{\rm dyn\,cm^{-2}}$ for SLy4 and BSk21, respectively. The near coincidence of the two curves is notable: once the shear modulus is constructed consistently from each EoS composition, the dependence of the axisymmetric stress amplitude on the choice between SLy4 and BSk21 is modest.

\section{Where the crust carries the load}
\begin{figure}
\centering
\includegraphics[width=\columnwidth]{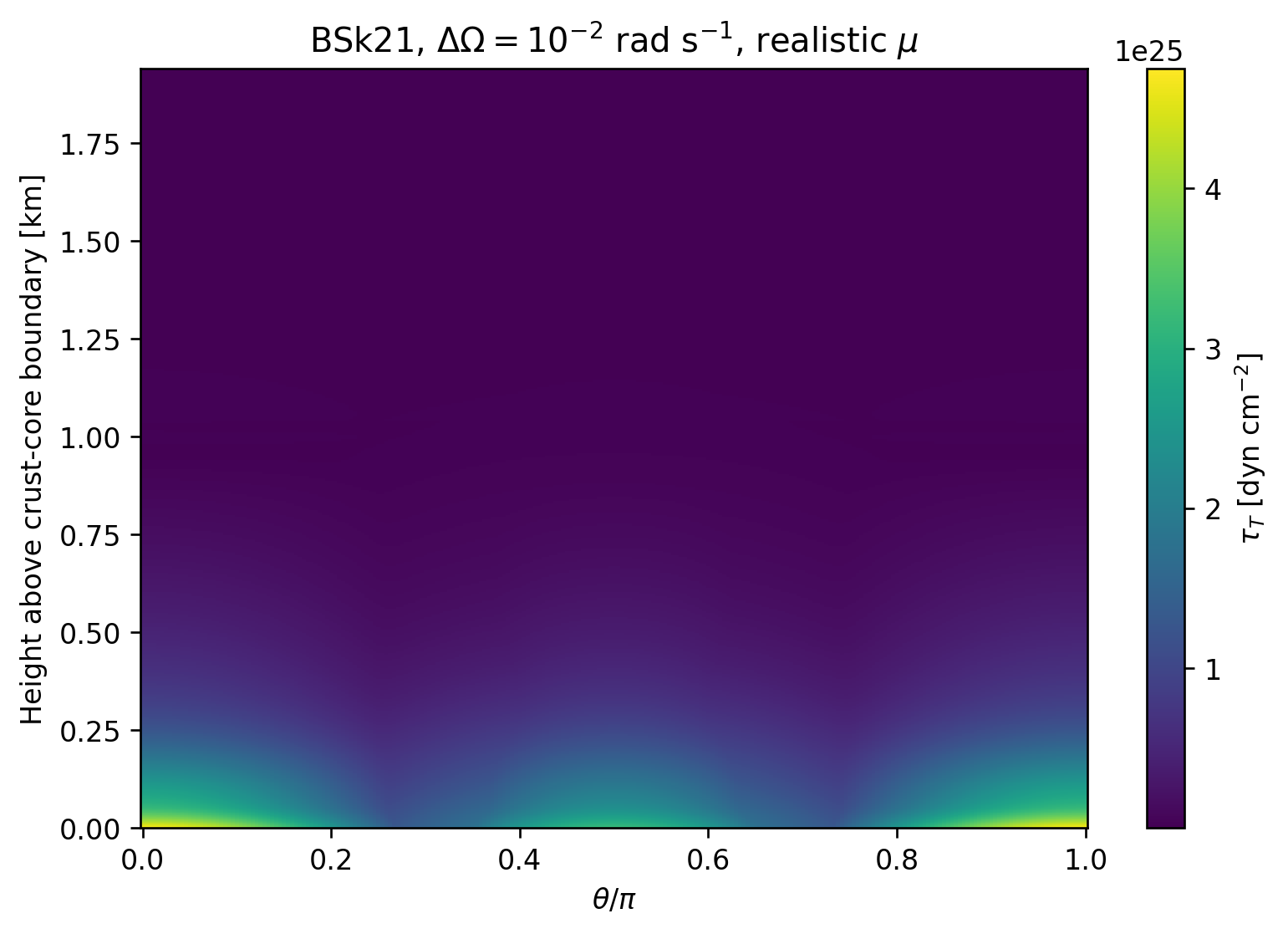}
\caption{Tresca stress in the BSk21 fiducial model with realistic Coulomb elasticity. Height is measured from the crust--core boundary. The maximum remains deep and polar even though the local Magnus force vanishes on the rotation axis.}
\label{fig:map}
\end{figure}

The most striking spatial result is that the maximum Tresca stress does not occur where the local Magnus force is largest. For both EoS and at every lag in the scan, the maximum lies on the crustal side of the crust--core boundary and towards the rotation axis (Fig.~\ref{fig:map}). Yet equation~(\ref{eq:magnus}) contains $\sin\theta$ and the local Magnus force vanishes on the axis.

The deep-polar maximum is therefore a property of the global elastic response, not a local image of the forcing. The pinned region loads a stratified shell that is mechanically coupled to the fluid core and gravitationally coupled to the whole star. Radial and tangential displacements redistribute the load before the stress invariant is formed. This is the same reason why the stress and strain maxima need not coincide: from equation~(\ref{eq:tresca}), a low-$\mu$ layer can develop a relatively large strain while carrying little shear stress. In the diagnostics used here the two quantities are tracked independently. Figure~3 displays the Tresca-stress field and establishes the deep-polar location of $\tau_{\rm T,max}$; the quoted $\alpha_{\max}$ values are failure diagnostics and should not be read as a statement that their spatial maxima are necessarily co-located with $\tau_{\rm T,max}$.

\section{Vela as an observational scale}
\subsection{Inter-glitch accumulation}
The fiducial lag is intended as a Vela-scale benchmark, not as a direct measurement of the internal differential rotation. Vela has a spin period of about $89$ ms and major glitches recur on a characteristic timescale of roughly three years \citep{Manchester2005,Espinoza2011}. With $|\dot\Omega|\simeq10^{-10}\,{\rm rad\,s^{-2}}$, a reservoir that does not follow the observed secular spin-down would accumulate
\begin{equation}
 \Delta\Omega_{\rm ig}\sim |\dot\Omega|t_{\rm ig}
 \sim (0.8-1.0)\times10^{-2}\,{\rm rad\,s^{-1}}
\end{equation}
for $t_{\rm ig}\sim2.5-3$ yr. The conversion of this accumulation scale into the actual local superfluid--crust lag is uncertain at order unity because vortex creep and motion, entrainment, coupling to the core and the previous glitch history all enter. Our lag scan spans $10^{-4}$--$1\,{\rm rad\,s^{-1}}$, so none of the qualitative conclusions depends on assigning high precision to the fiducial value.

At $\Delta\Omega=10^{-2}\,{\rm rad\,s^{-1}}$ the maximum strain is
\begin{align}
 \alpha_{\max}^{\rm SLy4}&=3.14\times10^{-5},\\
 \alpha_{\max}^{\rm BSk21}&=4.68\times10^{-5}.
\end{align}
These values are more than three orders of magnitude below the canonical breaking strain $\alpha_{\rm break}\sim0.1$ inferred from molecular-dynamics simulations \citep{HorowitzKadau2009}. Even the formal pinning-limited plateau gives only $9.7\times10^{-5}$ for SLy4 and $1.37\times10^{-4}$ for BSk21. Pinning alone therefore cannot fracture an initially relaxed crust.

\subsection{Angular-momentum sanity check}
The 2016 December Vela glitch had $\Delta\nu/\nu=1.431\times10^{-6}$ \citep{Palfreyman2018,GugercinogluAlpar2020}. With $\Omega\simeq70.6\,{\rm rad\,s^{-1}}$, this corresponds to
\begin{equation}
 \Delta\Omega_{\rm g}\simeq1.01\times10^{-4}\,{\rm rad\,s^{-1}}.
\end{equation}
A deliberately minimal angular-momentum budget gives
\begin{equation}
 I_{\rm s}\Delta\Omega_{\rm lag}\gtrsim I\Delta\Omega_{\rm g},
\end{equation}
if a reservoir of moment of inertia $I_{\rm s}$ efficiently releases a lag $\Delta\Omega_{\rm lag}$. For the fiducial $10^{-2}\,{\rm rad\,s^{-1}}$,
\begin{equation}
 \frac{I_{\rm s}}{I}\gtrsim1.0\times10^{-2},
\end{equation}
rising to $1.26$ per cent for $\Delta\Omega_{\rm lag}=8\times10^{-3}\,{\rm rad\,s^{-1}}$. For a canonical $I=10^{45}\,{\rm g\,cm^2}$ the 2016 event corresponds to $\Delta L\simeq1.0\times10^{41}\,{\rm g\,cm^2\,s^{-1}}$.

This estimate is intentionally not a glitch model: it assumes efficient release and says nothing about the spatial distribution of the reservoir. More importantly, the $1.0$--$1.26$ per cent values above are \emph{single-event minimum requirements} for an assumed lag; they are not predictions for the available reservoir fraction. They are slightly below the classic long-term Vela activity bound of at least $1.4$ per cent \citep{Link1999}. Hence the single-glitch sanity check is of the correct order of magnitude but does not by itself satisfy the cumulative activity constraint. Crustal entrainment strengthens, rather than removes, this tension by increasing the effective reservoir requirement and can motivate participation beyond a purely crust-confined reservoir \citep{Chamel2013,Newton2015}. The appropriate conclusion is therefore limited: a lag of order $10^{-2}\,{\rm rad\,s^{-1}}$ is compatible with the angular-momentum scale of an individual large Vela glitch, while a self-consistent account of the long-term activity requires the reservoir and coupling physics that are outside the present elastostatic model.

\section{Mass dependence}
\begin{figure}
\centering
\includegraphics[width=\columnwidth]{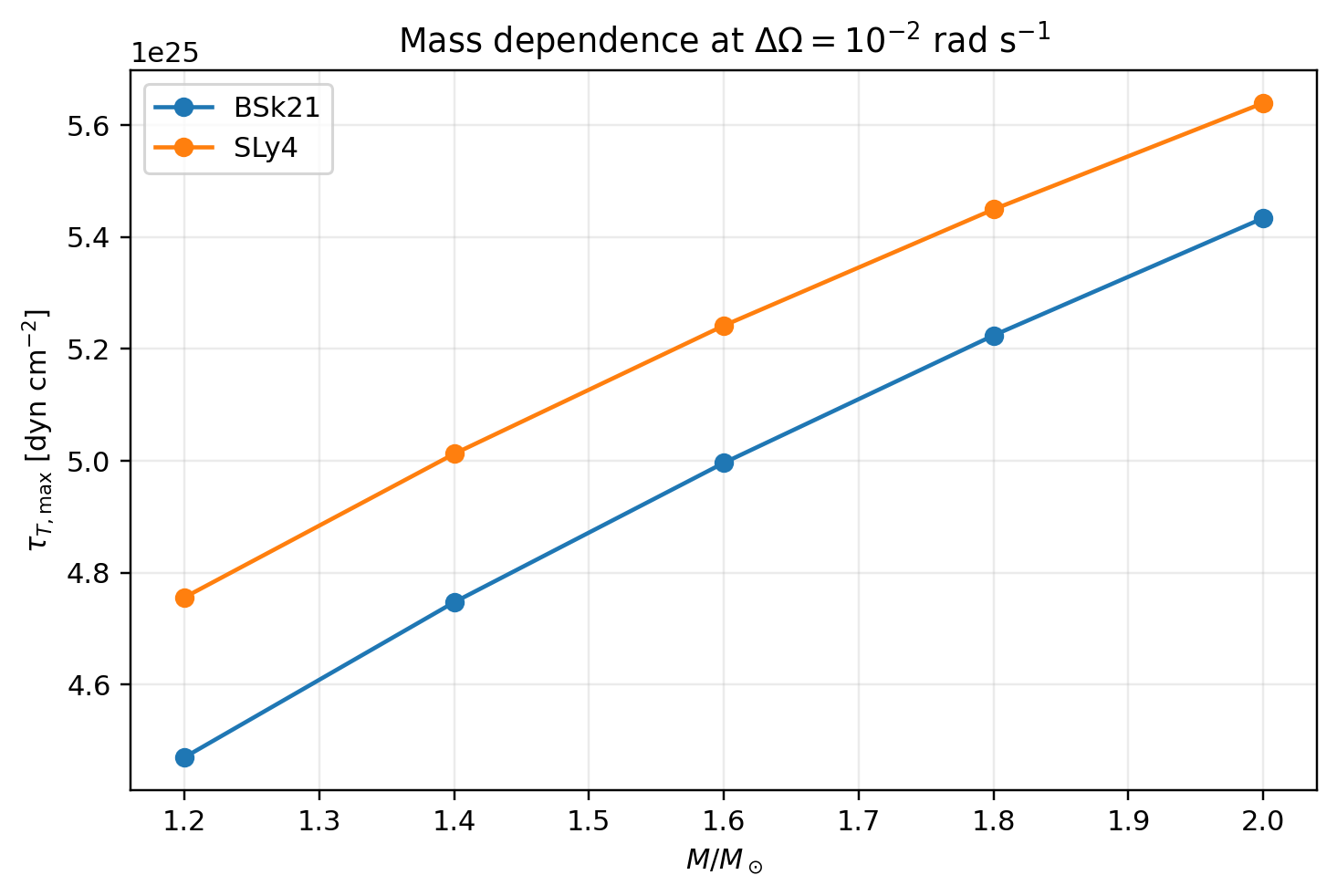}
\caption{Maximum Tresca stress at $\Delta\Omega=10^{-2}\,{\rm rad\,s^{-1}}$ for the realistic composition-dependent shear modulus. The mass dependence is smooth and modest for both EoSs, while the location of the maximum remains deep and polar throughout the scan.}
\label{fig:mass}
\end{figure}

To test whether the fiducial $1.4\,M_\odot$ model is representative, we repeat the realistic-$\mu$ calculation at $\Delta\Omega=10^{-2}\,{\rm rad\,s^{-1}}$ for $M=1.2$--$2.0\,M_\odot$. Figure~\ref{fig:mass} shows a smooth, monotonic but modest increase in $\tau_{\rm T,max}$. Between $1.2$ and $2.0\,M_\odot$ the increase is 18.6 per cent for SLy4 and 21.6 per cent for BSk21. At $2.0\,M_\odot$ the maximum strains are still only $3.56\times10^{-5}$ and $5.39\times10^{-5}$, respectively. In every model the stress maximum remains at the crust--core boundary towards the rotation axis. The mass therefore changes the amplitude at the tens-of-per-cent level but does not alter the qualitative mechanical regime or the failure conclusion.

\begin{table}
\centering
\caption{Mass scan at $\Delta\Omega=10^{-2}\,{\rm rad\,s^{-1}}$ with the composition-dependent Coulomb shear modulus. Stresses are in $10^{25}\,{\rm dyn\,cm^{-2}}$.}
\label{tab:mass}
\begin{tabular}{ccc}
\hline
$M/M_\odot$ & SLy4 & BSk21\\
\hline
1.2 & 4.756 & 4.470\\
1.4 & 5.012 & 4.747\\
1.6 & 5.242 & 4.996\\
1.8 & 5.450 & 5.224\\
2.0 & 5.640 & 5.434\\
\hline
\end{tabular}
\end{table}

\section{Discussion}
The calculation provides a direct mechanical map
\begin{equation}
 f_{\rm pin}(\rho)\rightarrow f_{\rm vol}(r,\theta)
 \rightarrow \boldsymbol\xi(r,\theta)
 \rightarrow \{\alpha,\tau_{\rm T}\}.
\end{equation}
Three aspects appear particularly robust. First, the low-lag linear response turns smoothly into a finite pinning-limited plateau because saturation occurs locally and progressively. Second, the deep-polar stress concentration is non-local and persists across the entire lag scan. Third, the induced strain remains far below the breaking scale even when the local pinning force is fully mobilized.

The realistic shear modulus strengthens these conclusions. In the controlled comparison of Appendix~\ref{app:numerics}, background, crust boundaries, pinning profile, angular projection and harmonic truncation are held fixed and only $\mu(r)$ is changed. Replacing the realistic modulus by $\mu=10^{-2}P$ lowers the SLy4 fiducial stress from $5.01$ to $4.54\times10^{25}\,{\rm dyn\,cm^{-2}}$ but raises the BSk21 value from $4.75$ to $5.62\times10^{25}\,{\rm dyn\,cm^{-2}}$. The EoS ordering is therefore reversed: realistic elasticity gives SLy4 slightly above BSk21, whereas $10^{-2}P$ gives BSk21 above SLy4. This sign reversal is more informative than either absolute correction. It shows that $10^{-2}P$ captures the qualitative mechanics but cannot be interpreted as an EoS-independent normalization of the realistic solution; the radial composition profile enters through the full elastic--gravitational operator.

Several limitations remain. The calculation is Newtonian and elastostatic. The superfluid is represented through a prescribed reaction force rather than an independent multifluid displacement, so entrainment and dynamical coupling are not solved self-consistently. The vortex array is coarse-grained and axisymmetric, and the crust is assumed initially relaxed and linearly elastic. These restrictions are deliberate: the present paper isolates the mechanical response to pinning before adding the extra physics required for non-axisymmetric mountains, history-dependent plasticity and glitch dynamics.

\section{Conclusions}
We have used the stratified elastic--gravitational framework of \citet{Giliberti2019,Giliberti2020} to calculate the axisymmetric crustal response to the reaction force of pinned superfluid vortices. The central result is not a particular stress normalization, but a robust mechanical picture.

At small lag the response is linear. As the Magnus force reaches the density-dependent mesoscopic cap, different parts of the crust saturate at different lags and the global stress turns smoothly towards a finite plateau. A Vela-scale $\Delta\Omega\sim10^{-2}\,{\rm rad\,s^{-1}}$ lies already in this partially saturated regime. Across both EoSs, the full lag scan and the $1.2$--$2.0\,M_\odot$ mass scan, the maximum stress remains at the deep crustal boundary towards the rotation axis even though the local Magnus force vanishes there. The localization is therefore a genuinely non-local consequence of elastic--gravitational load redistribution.

Composition-dependent elasticity is quantitatively important but does not change this picture. Relative to $\mu=10^{-2}P$, it changes the fiducial stress at the 10--20 per cent level and reverses the SLy4--BSk21 ordering. This demonstrates that simplified shear prescriptions cannot in general be absorbed into a universal multiplicative correction.

The induced strains remain very small: $\alpha_{\max}\sim3$--$5\times10^{-5}$ at the fiducial lag and $\lesssim1.4\times10^{-4}$ on the formal plateau, far below the canonical $\sim0.1$ breaking scale. The mass scan changes the stress by only about 20 per cent over $1.2$--$2.0\,M_\odot$ and does not modify this conclusion. Finally, the 2016 Vela glitch requires a minimum reservoir of order one per cent of the stellar moment of inertia if a lag of order $10^{-2}\,{\rm rad\,s^{-1}}$ is efficiently released, providing an independent angular-momentum sanity check on the fiducial scale.

Pinning therefore generates a structured, deep crustal pre-stress that is mechanically relevant but insufficient by itself to break an initially relaxed crust. This establishes the axisymmetric baseline needed before adding non-axisymmetric forcing, critical-lag/glitch dynamics and history-dependent plasticity in subsequent papers.
\section*{Data availability}
The numerical tables underlying the lag scan, mass scan, convergence tests and shear-modulus comparison are available from the author together with the scripts needed to reproduce the principal numerical results.

\appendix
\section{Angular projection of the cylindrical force}
\label{app:projection}
For $\boldsymbol f=f_{\rm vol}\boldsymbol e_\varpi$,
\begin{equation}
 f_r=f_{\rm vol}\sin\theta,\qquad f_\theta=f_{\rm vol}\cos\theta.
\end{equation}
Writing $x=\cos\theta$, the radial and spheroidal coefficients are
\begin{equation}
 h_R^{(\ell)}(r)=\frac{2\ell+1}{2}\int_{-1}^{1}f_r(r,x)P_\ell(x)\,dx,
\end{equation}
\begin{equation}
 h_S^{(\ell)}(r)=\frac{2\ell+1}{2\ell(\ell+1)}
 \int_{-1}^{1}f_\theta(r,x)\frac{dP_\ell}{d\theta}\,dx,
 \qquad \ell>0,
\end{equation}
with $h_S^{(0)}=0$. Equatorial symmetry leaves only even $\ell$. The local pinning cap is applied before this projection.

\section{Numerical and microphysical checks}
\label{app:numerics}
The production calculations use $\ell_{\max}=20$. Comparing $\ell_{\max}=16$ with 20 changes the fiducial stress by 0.022 per cent (SLy4) and 0.021 per cent (BSk21). The sharper high-lag forcing converges more slowly, but the corresponding plateau changes by only 0.75 and 0.92 per cent, respectively. The deep-polar location is unchanged. We therefore regard $\ell_{\max}=20$ as sufficient for the lag range and smooth endpoint prescription studied here. Sharper pinning endpoints or substantially more extreme high-lag configurations would require a renewed harmonic-convergence test and, if necessary, larger $\ell_{\max}$.

We also test radial integration convergence independently of the harmonic truncation. The perturbation equations are integrated with adaptive DOP853 and dense output; the reconstruction mesh is therefore not the integration mesh. At fixed $\ell_{\max}=8$, where the stress maximum is already at the crust--core boundary and pole, we halve the maximum allowed integration step from $(r_t-r_c)/120$ to $(r_t-r_c)/240$ and separately tighten the tolerances from $(r_{\rm tol},a_{\rm tol})=(5\times10^{-8},5\times10^{-10})$ to $(10^{-9},10^{-11})$. Relative to the tight solution, the nominal maximum stress differs by only $3.7\times10^{-6}$ for SLy4 and $1.34\times10^{-5}$ for BSk21; after halving the maximum step the differences are $6.6\times10^{-6}$ and $7.6\times10^{-6}$, respectively. The location of $\tau_{\rm T,max}$ is unchanged. Together with the independent $\ell_{\max}=16\rightarrow20$ test above, this shows that neither radial integration nor harmonic truncation controls the reported deep-boundary maximum.

The EoS-specific treatment of the inner endpoint of $f_{\rm pin}$ is also a small systematic. Retaining the original Seveso interpolation and truncating it at the crust--core boundary instead of bringing it continuously to zero changes the fiducial stress by 2.7 per cent for SLy4 and 1.5 per cent for BSk21, and the plateau by about one per cent.

Finally, we isolate the shear-modulus dependence by holding \emph{all other ingredients fixed}: the stellar background, EoS-specific crust boundaries, Seveso endpoint treatment, body-force normalization, angular projection, radial grid and $\ell_{\max}=20$. In this one-parameter comparison, replacing the realistic Coulomb modulus by $\mu=10^{-2}P$ gives $4.54\times10^{25}$ instead of $5.01\times10^{25}\,{\rm dyn\,cm^{-2}}$ for SLy4 and $5.62\times10^{25}$ instead of $4.75\times10^{25}\,{\rm dyn\,cm^{-2}}$ for BSk21 at $\Delta\Omega=10^{-2}\,{\rm rad\,s^{-1}}$. Hence the control law changes the realistic result by $-10.3$ per cent for SLy4 and $+18.5$ per cent for BSk21 and reverses the EoS ordering. Numbers obtained with a different crust-boundary/background realization are not part of this controlled comparison; throughout this paper, quoted production and control values refer to the common EoS-consistent background defined in Section~2. The opposite signs demonstrate that the realistic result cannot be obtained by a universal rescaling with $\mu$.

\label{lastpage}
\end{document}